\documentclass[preprintnumbers, onecolumn, floatfix, preprintnumbers, amsmath, amssymb, superscriptaddress, twocolumn]{revtex4-1}

\usepackage[colorlinks,linkcolor=red,urlcolor=blue,citecolor=blue]{hyperref}
\usepackage{amssymb, amsmath, bm, dcolumn, epsf, graphicx, latexsym, mathbbol, slashed}
\usepackage{mathrsfs}
\usepackage{comment}
\newcommand{\dd}[0]{\mathrm{d}}
\usepackage{graphicx}
\usepackage{epstopdf}
\usepackage{latexsym}
\usepackage{amssymb}
\usepackage{amsmath}
\usepackage{color}
\usepackage{mathrsfs}
\usepackage{xparse}
\usepackage{mathtools}
\usepackage{multirow}
\usepackage[center]{subfigure}
\usepackage{graphicx}
\usepackage[svgnames]{xcolor}
\usepackage{indentfirst}
\usepackage{titlesec}
\usepackage{fancyhdr,tabulary}
\usepackage{caption}
\usepackage{amsfonts,amsmath,amssymb,physics,amsthm}
\usepackage{booktabs,hyperref}
\usepackage{multirow}
\usepackage{verbatim}
\usepackage[section]{placeins}
\usepackage{float}

\begin{document}
\renewcommand\arraystretch{2}
 \newcommand{\bq}{\begin{equation}\begin{aligned}[b]}
 \newcommand{\eq}{\end{aligned}\end{equation}}
 \newcommand{\bqn}{\begin{eqnarray}}
 \newcommand{\eqn}{\end{eqnarray}}
 \newcommand{\nb}{\nonumber}
 \newcommand{\bvec}[1]{\mathbf #1}
 \newcommand{\cb}{\color{blue}}
    \newcommand{\cc}{\color{cyan}}
     \newcommand{\lb}{\label}
        \newcommand{\cm}{\color{magenta}}
\newcommand{\rc}{\rho^{\scriptscriptstyle{\mathrm{I}}}_c}
\newcommand{\rd}{\rho^{\scriptscriptstyle{\mathrm{II}}}_c}
\NewDocumentCommand{\evalat}{sO{\big}mm}{%
  \IfBooleanTF{#1}
   {\mleft. #3 \mright|_{#4}}
   {#3#2|_{#4}}%
}

\newcommand{\PRL}{Phys. Rev. Lett.}
\newcommand{\PL}{Phys. Lett.}
\newcommand{\PR}{Phys. Rev.}
\newcommand{\CQG}{Class. Quantum Grav.}
\newcommand{\parallelsum}{\mathbin{\!/\mkern-5mu/\!}}

\title{Electrodynamics with violations of Lorentz and U(1) gauge symmetries and their Hamiltonian structure}

\author{Xiu-Peng Yang${}^{a, b}$}
\email{youngxp@zjut.edu.cn}

\author{Bao-Fei Li${}^{a, b}$}
\email{libaofei@zjut.edu.cn}

\author{Tao Zhu${}^{a, b}$}
\email{corresponding author: zhut05@zjut.edu.cn}

\affiliation{
${}^{a}$ Institute for Theoretical Physics and Cosmology, Zhejiang University of Technology, Hangzhou, 310032, China\\
${}^{b}$ United Center for Gravitational Wave Physics (UCGWP), Zhejiang University of Technology, Hangzhou, 310032, China}

\begin{abstract}

This article aims to study the Lorentz/U(1) gauge symmetry-breaking electrodynamics in the framework of the Standard-Model Extension and analyze the Hamiltonian structure for the theory with a specific dimension $d\leq 4$ of Lorentz breaking operators. For this purpose, we consider a general quadratic action of the modified electrodynamics with Lorentz/gauge-breaking operators and calculate the number of independent components of the operators at different dimensions in gauge invariance and breaking. With this general action, we then analyze how the Lorentz/gauge symmetry-breaking can change the Hamiltonian structure of the theories by considering Lorentz/gauge-breaking operators with dimension $d\leq 4$ as examples. We show that the Lorentz-breaking operators with gauge invariance do not change the classes of the constraints of the theory and the number of the physical degrees of freedom of the standard Maxwell's electrodynamics. When the U(1) gauge symmetry-breaking operators are present, the theories in general lack first-class constraint and have one additional physical degree of freedom, compared to the standard Maxwell's electrodynamics. 

\end{abstract}

\maketitle

\section{Introduction}

The Standard Model (SM) successfully describes the fundamental constituents of matter using quarks, leptons, gauge bosons, and Higgs bosons. It effectively explains phenomena involving elementary particles and their interactions. The last predicted elementary particle of the SM, the Higgs boson, was experimentally confirmed in 2012 \cite{Higg1,Higg2}, giving a happy ending to the development of this great theory. General Relativity (GR) links gravity with the curvature of spacetime and provides a highly successful description of gravity-related phenomena. Since its inception, it has been recognized as a great theory and continues to be corroborated through various experiments. Its basic principles and predictions have been confirmed through multiple observations and measurements, establishing its foundational position in modern physics. In 2016, gravitational waves were observed in experiments \cite{GR}, which established the great position of general relativity. These two theories are expected to be unified at the Planck scale and potentially exhibit observable quantum gravity effects at accessible low-energy scales. This signal may be related to Lorentz symmetry breaking and can be described by an effective field theory \cite{EFT1}. 

To construct a consistent effective field theory that incorporates both GR and SM, Kosteleck\'y and \textcolor{black}{Colladay} proposed an effective field theory called the Standard Model Extension (SME) \cite{IS1,CV5}, which features general Lorentz and CPT violations. The measured and derived values of coefficients for Lorentz and CPT violations in the SME can be found in the data table organized by Kosteleck$\mathrm{\acute{y}}$ and  \textcolor{black}{Russell} \cite{IS4}. In recent years, research on the Cosmic Microwave Background (CMB) radiation \textcolor{black}{and the ultra high energy cosmic rays (UHECR) have} provided new opportunities for studying the pure photon sector of the SME. One reason is that any prediction involving the pure photon part that deviates from SM could potentially indicate Lorentz violation originating from the pure photon sector of the SME. A large amount of research has been conducted  \textcolor{black}{in this area \cite{CMB0, CMB1, CMB2, CMB3, CMB4, CMB5, CMB6, CMB7,UHE1,UHE2,UHE3,UHE4,UHE5,UHE6,UHE7,UHE8}}. Compared to conventional Maxwell electrodynamics, the pure-photon sector of the SME includes additional Lorentz-breaking terms, which can be classified as CPT-even and CPT-odd. The inclusion of these terms leads to the emergence of new effects, which has spurred extensive research in this field. \textcolor{black}{Studying the general aspects of the pure photon sector is a challenging task, Ref.~\cite{IS3} proposed a general electrodynamics extension theory with a quadratic action, which can be used to describe many related studies}, including photon interactions \cite{PSME1}, optical activity of media \cite{PSME2}, the Lorentz-invariance-Violating term \cite{HA1,HA2}, the Chern-Simons term \cite{PSME3}, nonminimal SME \cite{PSME4}, and other related phenomena involving photons \cite{PSME5}.  Theories with Lorentz-breaking operators of dimension 4 have received considerable attention, including the Lorentz-invariance-violating (LIV) \cite{KsLIV}, Carroll-Field-Jackiw (CFJ) \cite{TX1}, and Proca electrodynamics \cite{Proc}, with the first two being U(1) gauge invariant theories and the last one involving U(1) gauge symmetry breaking. In Refs. \cite{HA1,HA2,CFJH,HA3}, they conducted Hamiltonian analyses of the aforementioned three theories and identified the constraint structures of the theories, which motivates us to study the constraint structure of more general theories with a specific dimension $d\leq4$.

The Lagrangian density of quadratic electrodynamics, given in \cite{IS3}, is a quadratic polynomial in the photon field $A_{\alpha_1}$ and its higher-order derivatives $\partial_{\alpha_3}\ldots\partial_{\alpha_d}A_{\alpha_2}$ with $d\geq2$. \textcolor{black}{Since the constant coefficients $\mathcal{K}_{(d)}^{\alpha_{1} \alpha_{2} \alpha_{3} \ldots \alpha_{d}}$, which are contracted with $A_{\alpha_1}\partial_{\alpha_3}\ldots\partial_{\alpha_d}A_{\alpha_2}$, remain invariant under coordinate transformations, it leads to a violation of the Lorentz symmetry of the theory}. These constant coefficients can be regarded as originating from the vacuum expectation value of an operator in the underlying theory, or the dominant component of dynamic background fields, or an averaged effect. Through dimensional analysis, it can be found that the constants $\mathcal{K}_{(d)}^{\alpha_{1} \alpha_{2} \alpha_{3} \ldots \alpha_{d}}$ associated with $A_{\alpha_1}\partial_{\alpha_3}\ldots\partial_{\alpha_d}A_{\alpha_2}$ which is of dimension $d$ must have dimension $4-d$. Some researchers believe that the theories with power series that are renormalizable have mass dimension $d\leq 4$ \cite{Colladay:1998fq, IS3},  \textcolor{black}{and it is these theories with mass dimension $d\leq 4$ that are mainly studied in this work. Usually they are also those theories that contain only the first time derivative of the field $A_\mu$, which makes the theories healthy as this avoids the potential Ostrogradsky instability that the theory could possess} \cite{Svanberg:2022kdr, Woodard:2015zca}.

In this work, we will extend the U(1) gauge-invariant Lagrangian density of quadratic electrodynamics described in \cite{IS3} to the one that includes U(1) gauge symmetry breaking terms. We also perform Hamiltonian analysis for renormalizable specific dimension cases of $d\leq 4$. Our purpose is to clarify how the U(1) gauge breaking terms affect the constraint structure and physical degrees of freedom of the theory.

Our results indicate that the Lagrangian density of general quadratic electrodynamics is the combination of five terms. One corresponds to the Lagrangian density of standard electrodynamics, one is U(1) gauge invariant, one contains both  U(1) gauge invariant and breaking, and the remaining two are U(1) gauge breaking. The Hamiltonian analysis of the general gauge-breaking system with a specific dimension $d\leq4$ reveals that the system with $d=4 $ requires additional conditions to become a constrained system, while systems with $d=2 $ and $d=3 $ do not. It also shows that the Lorentz-breaking operators with gauge invariance do not change the classes of the constraints of the theory and the number of the physical degrees of freedom of the standard Maxwell's electrodynamics. When the U(1) gauge symmetry-breaking operators are presented, the theories in general lack first-class constraint and have one additional physical degree of freedom, compared to the standard Maxwell's electrodynamics.

The structure of this paper is as follows. The basic theory is discussed in Sec. \ref{s2}, where we give the number of independent components of the Lorentz-breaking operators with U(1) gauge violation and extend it to the general Lagrangian density containing U(1) gauge-breaking terms. Sec. \ref{se3} is about the Hamiltonian structure and degrees of freedom of theories with a specific dimension $d\leq4$. 
In Sec. \ref{se5}, We apply the obtained results to some specific models and derive the results for LIV, CFJ, and Proca electrodynamics. Our summary is in Sec. \ref{se6}.

 For the sake of clarity and conciseness, we will use two conventions: 
 \begin{enumerate}
     \item The Greek indices range from 0 to 3 and the  Latin indices run from 1 to 3. The metric of the background spacetime $\eta_{\mu\nu} \equiv(1, -1, -1, -1)$;
     \item The time argument of the vector field $A_\mu$ is suppressed throughout the manuscript, namely  $A_\mu(\bvec x) \equiv A_\mu(t,\bvec x)$.
 \end{enumerate}

\section{Electrodynamics with violations of Lorentz and U(1) gauge symmetry \label{s2}}

In this section, we present a brief introduction of the electrodynamics with quadratic action in the pure photon sector in the framework of SME with both Lorentz and U(1) gauge symmetry breaking. This represents an extension of the Lorentz-violating modified electrodynamics with U(1) gauge invariance in \cite{IS3} by including the U(1) gauge symmetry-breaking operators in the quadratic action. In the construction of this theory, we also analyze the properties of the coefficients for Lorentz- and U(1) gauge-violating operators.

For our purpose, following the similar construction performed in \cite{IS3}, let us start with the general quadratic action for the Lorentz-violating electrodynamics in the pure photon sector, which can be written as \cite{IS3}
\textcolor{black}{
\bqn
S = \int d^4 x {\cal L} \label{action}
\eqn
}
with
\bqn
\mathcal L=-\frac{1}{4}F_{\mu\nu}F^{\mu\nu}+\sum\limits_{d=2}^\infty \mathcal{K}_{(d)}^{\alpha_{1} \alpha_{2} \alpha_{3} \ldots \alpha_{d}} A_{\alpha_{1}} \partial_{\alpha_{3}} \ldots \partial_{\alpha_{d}} A_{\alpha_{2}}, \nonumber \\\label{eq1}
\eqn
where $F_{\mu\nu}\equiv\partial_\mu A_\nu-\partial_\nu A_\mu$ and $\mathcal{K}_{(d)}^{\alpha_{1} \alpha_{2} \alpha_{3} \ldots \alpha_{d}}$ are constant coefficients with mass dimension $4-d$. One possible explanation for the coefficients $\mathcal{K}_{(d)}^{\alpha_{1} \alpha_{2} \alpha_{3} \ldots \alpha_{d}}$ originates from non-zero vacuum expectation values to the Lorentz-tensor fields. Note that each term associated with the coefficient $\mathcal{K}_{(d)}^{\alpha_{1} \alpha_{2} \alpha_{3} \ldots \alpha_{d}}$ violates CPT if $d$ is odd or preserves CPT if $d$ is even.

The symmetry among indices  $\{\alpha_3,\ldots,\alpha_{d}\}$ of tensor $\partial_{\alpha_3}\ldots\partial_{\alpha_{d}}$ and the use of integration by parts results in two properties of the coefficients $\mathcal{K}_{(d)}^{\alpha_{1} \alpha_{2} \alpha_{3} \ldots \alpha_{d}}$. One is the total symmetry in the $d-2$ indices $\{\alpha_3,\cdots,\alpha_d\}$, another is the symmetry of the two indices $\{\alpha_1,\alpha_2\}$ when $d$ is even and antisymmetry when $d$ is odd. Depending on the specific intrinsic symmetries of the Lorentz-violating operators, the coefficients $\mathcal{K}_{(d)}^{\alpha_{1} \alpha_{2} \alpha_{3} \ldots \alpha_{d}}$ can be decomposed into five representations \cite{IS3}. When one imposes the conditions of the $U(1)$ gauge invariance on the Lorentz-violating operators, these five representations are reduced into two representations, one corresponds to the CPT-even coefficient and the other corresponds to CPT-odd coefficient \cite{IS3}. 

\subsection{Lorentz-violating electrodynamics with $U(1)$ gauge invariance}

Let us first consider the Lorentz-violating electrodynamics with $U(1)$ gauge invariance \cite{IS3}. The $U(1)$ gauge invariance is a symmetry of the theory under the $U(1)$ gauge transformation
\bqn
A_\mu \to \tilde{A}_\mu = A_\mu + \partial_\mu \Lambda,
\eqn
where $\Lambda$ is an arbitrary function. The variation of the action (\ref{action}) under this gauge transformation reads
\textcolor{black}
{
\bqn
\delta_g S&=&-\sum_{d=2}^{\infty}\int\dd^4 x\mathcal K_{(d)}^{\alpha_{1} \alpha_{2} \alpha_{3} \cdots \alpha_{d}} \Lambda\nonumber \\
&& \times\partial_{\alpha_3}\cdots\partial_{\alpha_d}\qty(\partial_{[\alpha_1} A_{\alpha_2]{\pm}}+\frac{1}{2}\partial_{[\alpha_1}\partial_{\alpha_2]\pm}\Lambda)=0,\nonumber\\
\label{equ7}
\eqn
}
\textcolor{black}{where $``+/-"$ corresponds to even/odd dimension, and the brackets $[]_+$ and $[]_-$ indicate symmetrization and antisymmetrization respectively}. Specifically, the terms in the big bracket of the above expression can be written as
\bqn
&&\partial_{[\alpha_1} A_{\alpha_2]{\pm}}+\frac{1}{2}\partial_{[\alpha_1}\partial_{\alpha_2]\pm}\Lambda \nb\\
&&~~~~=
\begin{cases}
\partial_{\alpha_1} A_{\alpha_2}+\partial_{\alpha_2} A_{\alpha_1}+\partial_{\alpha_1} \partial_{\alpha_2} \Lambda, & d\; \mathrm{is \;even},\\
\partial_{\alpha_1}  A_{\alpha_2}- \partial_{\alpha_2} A_{\alpha_1}, & d \; \mathrm{is\;odd}.
\end{cases}
\label{eq2.9}
\eqn
The $U(1)$ gauge invariance requires the variation of the action in (\ref{equ7}) vanishes. When $d$ is even, since the two first indices $\alpha_1$ and $\alpha_2$ are symmetric, in order to make (\ref{equ7}) vanishes, one has to require both indices $\alpha_1$ and $\alpha_2$ are antisymmetric with one of $\{\alpha_3, \alpha_4, \cdots, \alpha_d\}$ \cite{IS3}. With these properties, it is straightforward to infer that all the CPT-even operators with $d=2$ are gauge-violating and the gauge invariance requires $d\geq 4$. By using these properties, one can also count the number of independent components of the CPT-even operators, which leads to
\bqn
N_F = (d+1)d(d-3).
\eqn
When $d$ is odd, similarly, the first two indices $\alpha_1$ and $\alpha_2$ are antisymmetric, and the vanishing of (\ref{equ7}) require $\alpha_1$ and $\alpha_2$ are antisymmetric with one of $\{\alpha_3, \alpha_4, \cdots, \alpha_d\}$. For this case, only operators with $d\geq 3$ are allowed and the number of independent components of the CPT-odd operators is
\bqn
N_{AF} = \frac{1}{2}(d+1)(d-1)(d-2).
\eqn
Note that in Table.~\ref{tb1}, we summarize the number of independent components for the CPT-odd and CPT-even operators.

\subsection{Lorentz-violating electromagnetics with the violations of $U(1)$ gauge symmetry}

Now let us turn to consider the case with the breaking of the $U(1)$ gauge symmetry. When the $U(1)$ gauge symmetry is violated, one does not require the variation of the action in (\ref{equ7}) vanishing. For this case, one does not need to impose extra conditions on the coefficients $\mathcal{K}_{(d)}^{\alpha_{1} \alpha_{2} \alpha_{3} \ldots \alpha_{d}}$ to ensure the gauge invariance of the theory. For CPT-even operators without gauge invariance, as we mentioned before, while the indices $\{\alpha_3, \alpha_4, \cdots, \alpha_d\}$ in the CPT-even coefficients $\mathcal{K}_{(d)}^{\alpha_{1} \alpha_{2} \alpha_{3} \ldots \alpha_{d}}$ are symmetric, first two indices $\alpha_1$ and $\alpha_2$ are also symmetric. Similarly, for CPT-odd operators without gauge invariance, the indices $\{\alpha_3, \alpha_4, \cdots, \alpha_d\}$ are symmetric, while the first two indices $\alpha_1$ and $\alpha_2$ are antisymmetric. One can then count the  number of independent components of the CPT-even and CPT-odd operators with a specific dimension $d$, which are summarized in Table.~\ref{tb1}.

\begin{table}[H]
\caption{\label{tb1}%
The number of independent components of the Lorentz-violating operators with/without $U (1)$ gauge invariance.}
\begin{ruledtabular}
\centering
\begin{tabular}{cccc}
\multirow{2}{1cm}{\centering $d$} & \multirow{2}{1cm}{CPT} & Gauge & Number of \\
&~ & invariance & independent components \\
\hline
2&even&no&10\\
\hline
\multirow{2}{1cm}{\centering even, $\geq4$}&\multirow{2}{1cm}{even}&yes&$(d+1)d(d-3)$\\
	&&no&$\frac{2}{3} (d+2)(d+1)d$\\
 \hline
\multirow{2}{1cm}{\centering odd, $\geq3$}	&\multirow{2}{1cm}{odd}&yes&$\frac{1}{2}(d+1)(d-1)(d-2)$\\
	&&no&$\frac{1}{2}(d+2)(d+1)(d-1)$
\end{tabular}
\end{ruledtabular}
\end{table}

For later convenience, one can decompose the Lagrangian density (\ref{eq1}) into two parts, the $U(1)$ gauge invariance part and the gauge-violating part. Specifically, this involves rewriting the Lorentz-breaking coefficients in five different forms with distinct symmetries. The specific representation decomposition of these five forms can be found in \cite{IS3}. Then, the Lagrangian density (\ref{eq1}) can be rewritten as
\bqn
\mathcal L&=&-\frac{1}{4}F_{\alpha\nu}F^{\alpha\nu}  \nonumber \\
&& + \sum_{{\rm even}\; d=2}^\infty\mathcal{K}_{(d)}^{(1)\alpha_1\alpha_2\alpha_3\ldots\alpha_{d}} A_{\alpha_1}\partial_{\alpha_3}\ldots\partial_{\alpha_{d}} A_{\alpha_2}\nonumber\\
&& +\sum_{d=3}^\infty\mathcal{K}_{(d)}^{(2)\alpha_1\mu\nu\alpha_2\ldots\alpha_{d-2}}A_{\alpha_1}\partial_\nu\partial_{\alpha_2}\ldots\partial_{\alpha_{d-2}} A_\mu \nonumber \\
&& +\sum_{d=3}^\infty\mathcal{K}_{(d)}^{(3)\mu\alpha_1\nu\alpha_2\ldots\alpha_{d-2}}A_{\mu}\partial_\nu\partial_{\alpha_2}\ldots\partial_{\alpha_{d-2}} A_{\alpha_1}\nonumber\\
&&+ \sum_{d=4}^\infty\mathcal{K}_{(d)}^{(4)\mu\nu\rho\sigma\alpha_{1}  \ldots \alpha_{d-4}}A_{\mu} \partial_{\rho} \partial_{\sigma}\partial_{\alpha_1}\ldots \partial_{\alpha_{d-4}} A_{\nu} \nonumber \\
&& +\sum_{{\rm odd}\; d=3}^\infty\mathcal{K}_{(d)}^{(5)\mu\nu\rho\alpha_1\ldots\alpha_{d-3}} A_{\mu}\partial_{\rho}\partial_{\alpha_1}  \ldots\partial_{\alpha_{d-3}} A_{\nu}, \nonumber\\
 \label{eq2.03}
\eqn
where the five coefficients ${\cal K}_{(d)}^{(i)\alpha_1\alpha_2 \alpha_3 \cdots \alpha_d}$ with $i=1, 2, 3, 4, 5$ are distinguished by their distinct symmetries among the indices $\{\alpha_1, \alpha_2, \alpha_3, \cdots, \alpha_d\}$. The coefficients ${\cal K}_{(d)}^{(1)\alpha_1\alpha_2 \alpha_3 \cdots \alpha_d}$ are all CPT-even coefficients with $d\geq 2$, while their indices $\{\alpha_1, \alpha_2, \alpha_3, \cdots, \alpha_d\}$ are all symmetric. It is obvious to infer from (\ref{equ7}) that these coefficients violate the $U(1)$ gauge symmetry of the theory. For the coefficients ${\cal K}_{(d)}^{(2)\alpha_1 \mu \nu\alpha_2 \cdots \alpha_{d-2}}$, they are be either CPT-even or CPT-odd with $d \geq 3$. Except the two indices $\mu$ and $\nu$ are antisymmetric, the rest indices $\{\alpha_1, \alpha_2, \cdots, \alpha_{d-2}\}$ are all symmetric. Note that these coefficients are also gauge-violating. Similarly, for ${\cal K}_{(d)}^{(3) \mu\alpha_1 \nu\alpha_2 \cdots \alpha_{d-2}}$, they can also be either CPT-even or CPT-odd with $d \geq 3$. The two indices of this coefficient $\mu$ and $\nu$ are antisymmetric and the rest indices $\{\alpha_1, \alpha_2, \cdots, \alpha_{d-2}\}$ are all symmetric. One can check that these coefficients also break the $U(1)$ gauge symmetry of the theory. Then for the indices in the coefficients ${\cal K}_{(d)}^{(4) \mu\nu \rho\sigma\alpha_1\ldots \alpha_{d-4}}$, the two indices $\mu$ and $\rho$, and the two indices $\nu$ and $\sigma$, are both antisymmetric. We also note that these coefficients are symmetric upon interchange of the two pairs of indices $(\mu, \rho)$ and $(\nu, \sigma)$. By inspecting the variation of the action (\ref{equ7}) with these coefficients, one can infer that the CPT-odd operators with ${\cal K}_{(d)}^{(4) \mu\nu\rho\alpha_1 \cdots \alpha_{d-4}}$ violates the gauge invariance, while the CPT-even ones are gauge invariance. The last coefficients $\mathcal{K}_{(d)}^{(5)\mu\nu\rho\alpha_1\ldots\alpha_{d-3}}$ are CPT-odd coefficient with $d\geq 3$, and their three indices $\{\mu, \nu, \rho\}$ are antisymmetric. These coefficients preserve the $U(1)$ gauge symmetry of the theory. 

To simplify the later handling of the Hamiltonian analysis of the theory, one can rewrite the Lagrange density of the theory in a compact form of
\bqn\label{eq2.09}
 \mathcal L&=&-\frac{1}{4}F_{\alpha\nu}F^{\alpha\nu}+A_{\alpha_1}\hat{\mathcal{K}}^{(1)\alpha_1\alpha_2}A_{\alpha_2}+\frac{1}{2}A_{\alpha_1}\hat{\mathcal{K}}^{(2,3)\mu\nu\alpha_1} F_{\nu\mu}\nonumber\\
 && -\frac{1}{4}F_{\mu\rho}\hat{\mathcal{K}}^{(4)\mu\rho\nu\sigma}F_{\nu\sigma}+\frac{1}{2}\epsilon^{\kappa\mu\nu\rho}A_{\mu}\hat{\mathcal{K}}^{(5)}_{\kappa} F_{\nu\rho},
 \label{eq2.06}
\eqn
where
\bqn
\hat{\mathcal{K}}^{(1)\alpha_1\alpha_2}&\equiv& \sum_{d=\mathrm{even}}\mathcal{K}_{(d)}^{(1)\alpha_1\alpha_2\alpha_3\ldots\alpha_{d}} \partial_{\alpha_3}\ldots\partial_{\alpha_{d}}, 
\eqn
\bqn
\hat{\mathcal{K}}^{(4)\mu\rho\nu\sigma}&\equiv& \sum_{d=4}^\infty\mathcal{K}_{(d)}^{(4)\mu\nu\rho\sigma\alpha_{1}  \ldots \alpha_{d-4}}\partial_{\alpha_1}\ldots \partial_{\alpha_{d-4}},
\eqn
\bqn
\hat{\mathcal{K}}^{(5)}_{\kappa} &\equiv & \frac{\epsilon_{\kappa\mu\nu\rho}}{6}\sum_{d=\mathrm{odd}}\mathcal{K}_{(d)}^{(5)\mu\nu\rho\alpha_1\ldots\alpha_{d-3}} \partial_{\alpha_1}\ldots\partial_{\alpha_{d-3}},\nb\\
\eqn
and
\bqn
&&\hat{\mathcal{K}}^{(2,3)\mu\nu\alpha_1}\equiv \nb\\
&&~~~~~~~~ \sum_{d=3}^\infty \left[\mathcal{K}_{(d)}^{(2)\alpha_1\mu\nu\alpha_2\ldots\alpha_{d-2}} +(-1)^d\mathcal{K}_{(d)}^{(3)\mu\alpha_1\nu\alpha_2\ldots\alpha_{d-2}}\right] 
\nb \\
&& ~~~~~~~~~~~~ \times \partial_{\alpha_2}\ldots\partial_{\alpha_{d-2}}.
\eqn
The indice symmetries of the operators in (\ref{eq2.09}) are shown in Table. \ref{mlt3}. The indices enclosed in the same braces $\{\}$ of the second and third columns represent symmetry and antisymmetry between them, respectively. $\{\mu\rho,\nu\sigma\}$ in the second column indicates that the corresponding operators are symmetry when the two pairs $(\mu,\rho)$ and $(\nu,\sigma)$ are exchanged. The fourth column displays the conditions under which each class of operators appears.

\begin{table}[htb]
\caption{ Symmetries of the indices of the coefficients of operators in (\ref{eq2.09}).}
\begin{ruledtabular}
\centering
\begin{tabular}{cccc}
Coefficient &Symmetry&Antisymmetry&$d$\\
\hline
$\hat{\mathcal K}^{(1)\alpha_1\alpha_2}$&$\{\alpha_1,\alpha_{2}\}$&$\cdots$&even,$\;\geq2$\\
$\hat{\mathcal K}^{(2,3)\mu\nu\alpha_1}$&$\cdots$&$\{\mu,\nu\}$&$\geq3$\\
$\hat{\mathcal K}^{(4)\mu\rho\nu\sigma}$&$\{\mu\rho,\nu\sigma\}$&$\{\mu,\rho\},\{\nu,\sigma\}$&$\geq 4$\\
$\hat{\mathcal K}_{\kappa}^{(5)}$&$\cdots$&$\cdots$&odd,$\;\geq3$\\
\end{tabular}
\end{ruledtabular}
\label{mlt3}
\end{table}

\section{Hamiltonian structure of the theory\label{se3}}

In this section, we perform Hamiltonian analysis on the Lorentz-violating electromagnetic with and without $U(1)$ gauge invariance by using the Dirac-Bergmann procedure \cite{Berg1,Berg2,Dirac1,Dirac2,Dirac3}. For simplicity, we will focus on the theory with Lagrangian densities with a specific dimension $d\leq 4$ of Lorentz/gauge-breaking operators.

\subsection{$d=4$}

We start the Hamiltonian analysis for $d=4$. For this case, the Lagrangian density of the theory reduces to
\bqn\label{eqL4}
{\cal L}_{(4)} &=&-\frac{1}{4}F_{\mu\nu}F^{\mu\nu}-\frac{1}{2}{U}^{\mu\nu\rho\sigma}\partial_{\mu}A_\nu\partial_{\rho} A_{\sigma} \nb \\
&& -\frac{1}{2}V^{\mu\nu\rho\sigma}F_{\nu\mu}\partial_{\sigma}A_{\rho}-\frac{1}{4}W^{\mu\rho\nu\sigma}F_{\mu\rho}F_{\nu\sigma},\label{L4}
\eqn
where
\bqn
U^{\mu\nu\rho\sigma}&=&2\mathcal{K}_{(4)}^{(1)\mu\nu\rho\sigma},\\
V^{\mu\nu\rho\sigma}&=&\mathcal{K}_{(4)}^{(2)\rho\mu\nu\sigma}+\mathcal{K}_{(4)}^{(3)\mu\rho\nu\sigma},\\
W^{\mu\rho\nu\sigma}&=&\mathcal{K}_{(4)}^{(4)\mu\nu\rho\sigma}.
\eqn
The terms in ${\cal L}_{(4)}$ with coefficients $U^{\mu\nu\rho\sigma}$ and $V^{\mu\nu\rho\sigma}$ break the U(1) gauge symmetry. The four indices $\{\mu, \nu, \rho, \sigma\}$ in $U^{\mu\nu\rho \sigma}$ are total symmetric, while $V^{\mu\nu\rho\sigma}$ are antisymmetric in $\{\mu,\nu\}$ and symmetric in $\{\rho,\sigma\}$. The terms with coefficients $W^{\mu\rho\nu\sigma}$ are gauge invariant and the corresponding indices of the coefficients $W^{\mu\rho\nu\sigma}$ have the same symmetric properties of the Riemann tensor, i.e. the first pair $\{\mu\rho\}$ and the last pair $\{\nu\sigma\}$ of $W^{\mu\rho\nu\sigma}$ are both antisymmetric, and symmetric upon interchange of the two pairs.  
Variation the action (\ref{action}) with the above Lagrangian with respect to the field $A_\mu$, one obtains the equation of motion of the electromagnetic, i.e., 
\bqn
&&\partial_\mu\Big(F^{\mu\nu}+U^{\mu\nu\rho\sigma}\partial_\rho A_\sigma+V^{\nu\mu\rho\sigma}\partial_{\sigma}A_{\rho} \nb\\
&&~~~~~~~~ +\frac{1} {2}V^{\rho\sigma\nu\mu} F_{\sigma\rho}+W^{\rho\sigma\mu\nu}F_{\rho\sigma}\Big)=0.
\eqn

Now to perform the Hamiltonian analysis, it is convenient to define the conjugate momentum
\bqn
\pi^\mu\equiv\pdv{\mathcal L_{(4)}}{\dot A_\mu}&=&-F^{0\mu}-U^{0\mu\nu\rho}\partial_\nu A_\rho-V^{\mu0\nu\rho}\partial_{\rho}A_{\nu} \nb\\
&& -\frac{1}{2}V^{\nu\rho\mu0}F_{\rho\nu}-W^{0\mu\nu\rho}F_{\nu\rho},
\label{eq12}
\eqn
which makes the fundamental Poisson brackets (PB) as
\bqn
\{A_\mu(\bvec x),\pi^\nu(\bvec y)\}=\delta^\nu_\mu\delta(\bvec x-\bvec y).
\eqn
From the conjugate momentum, the canonical Hamiltonian of the system can be expressed as
\bqn
{\cal H}_{(4)} \equiv \dot{A}_{\mu} \pi^\mu - {\cal L}_{(4)}.
\eqn
In Hamiltonian analysis, a significant portion of the work involves computing the Poisson bracket between the Hamiltonian and functions in phase space. In this context, it is important to express the Hamiltonian as a function of the conjugate momenta and coordinates. To do so, let's write the time and spatial components of the conjugate momentum for (\ref{eq12}),
\bqn
\pi^0&=&-U^{00\mu\nu}\partial_\mu A_\nu-\frac{1}{2}V^{\mu\nu00}F_{\nu\mu},\\
\pi^k&=&{D^{k}}_i F^{0i}+N^k,
\label{eq14}
\eqn
where the matrices ${D^k}_i$ and $N^k$ are defined as
\bqn
{D^{k}}_i & \equiv & -{\delta^{k}}_i-{V_{i0}}^{ k0}-{V^{k0}}_{0i}-2{W^{0k}}_{0i}-{U^{0k}}_{0i},\\
N^k &\equiv& -\Big(\frac{1}{2} V^{ji k0}+W^{0kij}\Big)F_{ij}-(U^{0k00}+V^{k000})\partial_{0}A_{0}\nonumber\\
&&-2(U^{0k0i}+V^{k00i})\partial_i A_0-(U^{0kij}+V^{k0ij})\partial_i A_j.\nb\\
\eqn
From the expression of ${D^k}_i$, it is easy to get that ${D^k}_i = {D_i}^k$. Further analysis is required to determine the constraint structure of the system. For this purpose, we assume that ${D^k}_i$ is a non-degenerate matrix, just like in gauge invariance \cite{HA1,HA2}, such that
\bqn
    F^{0i}=&{(D^{-1})^{i}}_{k}(\pi^k-N^k).
\eqn 
After some manipulations, one can finally express the canonical Hamiltonian density as a function of the field quantity and its conjugate momentum, namely
\begin{widetext}
\bqn
\mathcal H_{(4)}&=&\pi^k\partial_kA_0+\frac{1}{2}{(D^{-1})_{jl}}(\pi^l-N^l)\Big[\pi^j-N^j+({V^{0j}}_{0k}-{V_{k0}}^{j0}){(D^{-1})^{k}}_{i}(\pi^i-N^i)-2(U^{0j00}+V^{j000})\partial_0A_0\Big]\nonumber\\
&&+\frac{1}{2}F_{ij}\qty(\frac{1}{2}F^{ij}+\frac{1}{2}W^{ijkl}F_{kl}+2V^{ji0k}\partial_{k}A_{0}+V^{jikl}\partial_{l}A_{k})+\qty(\frac{1}{2}U^{ijkl}\partial_i A_j+2U^{0jkl}\partial_j A_0)\partial_k A_l\nonumber\\
&&+2U^{00kl}\partial_k A_0\partial_l A_0-\frac{1}{2}U^{0000}\partial_0A_0\partial_0 A_0,
\label{eq17}
\eqn
\end{widetext}
\textcolor{black}{where we keep the term of $\partial_0 A_0$ explicitly for the reason that when studying the constrained system with $d=4$ in the following we have to impose some constraints on the Lorentz-breaking coefficients and one of our choice here is exactly $U^{0000}=0$ that can be seen as throwing away the term with respect to  $\partial_0 A_0$}.

Following the Dirac-Bergmann algorithm, our next step is to write the total Hamiltonian density of the system, which is composed of the canonical Hamiltonian density and primary constraints. However, we will see that for the system in $d=4$, the existence of primary constraints requires certain additional conditions.

\subsubsection{Conditions for the existence of the constraints}

Let us look at what conditions the system described by (\ref{L4}) contains constraints. According to the Dirac-Bergmann procedure, the presence of constraints in the system with Lagrangian density $\mathcal L_{(4)}$ can be determined by checking whether the Hessian matrix $\pdv[2]{\mathcal L}{\dot A_\mu}{\dot A_\nu}$ is degenerate. Therefore, for making the system with Lagrangian density $\mathcal L_{(4)}$ a constrained system, it is sufficient to provide the condition that the matrix $\pdv[2]{\mathcal L_{(4)}}{\dot A_\mu}{\dot A_\nu}$ is degenerate. The Hessian matrix takes the form,
\bqn\label{eq28}
\pdv[2]{\mathcal L_{(4)}}{\dot A_\mu}{\dot A_\nu} &=& -\qty(\eta^{\mu\nu}-\eta^{0\mu}{\delta_0}^{\nu})-2W^{0\mu0\nu}-U^{0\mu0\nu} \nb\\
&& -V^{\mu0\nu0}-V^{\nu0\mu0}.
\eqn
When the Hessian matrix is non-degenerate, the system does not possess any constraint that is not of our interest. It is easy to observe that for gauge invariant case since one has
\bqn
U^{\mu\nu\rho\sigma}=0=V^{\mu\nu\rho\sigma},
\eqn
the Hessian matrix is identically degenerate. This indicates that the theory with $U(1)$ gauge symmetry always has constraints. However, when the $U(1)$ gauge symmetry is broken, whether the theory possesses constraints depends on the specific forms of the gauge-breaking coefficients $U^{\mu\nu\rho\sigma}$ and $V^{\mu\nu\rho\sigma}$.

For our purpose, we intend to consider the case when the Hessian matrix is degenerate, either with gauge invariance or gauge breaking.  Considering the complexity of the Lagrangian density for $d=4$ case, for simplicity, let us only focus on the following simple case to make the Hessian matrix degenerate, 
\bqn
U^{0000}=0,\quad U^{000i}=-V^{i000}.
\label{eq4.8}
\eqn
From the above equation, it can be seen that this constraint condition only restricts the gauge-breaking coefficients and the first row and column of Hessian matrix that are 0, which allows the subsequent conclusions to be fully applicable to gauge-invariant systems. For the discussion clarity, similar to gauge invariance case \cite{HA1,HA2}, we will assume that the $3\times 3$ matrix $\pdv[2]{\mathcal L_{(4)}}{\dot A_i}{\dot A_j}$ is non-degenerate.  

Thus, for $d=4$ case, hereafter we will analyze the system with 
\bqn
\mathcal L_{(4)}&=&-\frac{1}{4}F_{\mu\nu}F^{\mu\nu}-\frac{1}{2}{U}^{\mu\nu\rho\sigma}\partial_{\mu}A_\nu\partial_{\rho} A_{\sigma} \nb\\
&&-\frac{1}{2}V^{\mu\nu\rho\sigma}F_{\nu\mu}\partial_{\sigma}A_{\rho}-\frac{1}{4}W^{\mu\rho\nu\sigma}F_{\mu\rho}F_{\nu\sigma},
\label{eq27}
\eqn
with conditions $U^{0000}=0$ and $U^{000i}=-V^{i000}$. 

\subsubsection{Canonical analysis\label{se2.2}}

After obtaining the constrained system with a Lagrangian density of (\ref{eq27}), this subsection analyzes the constraint structure of the system. Under the condition of (\ref{eq4.8}), the conjugate momentum in (\ref{eq14}) can be written as 
\bqn
\pi^0&=&-(U^{00ij}+V^{ji00})\partial_i A_j-2U^{000i}\partial_iA_0, \\
\pi^k&=&{D^{k}}_i F^{0i}+N^k,
\label{eq48}
\eqn
where 
\bqn
N^k&=&-(\frac{1}{2}V^{ji k0}+W^{0kij})F_{ij}-2(U^{0k0i}+V^{k00i})\partial_i A_0 \nb\\
&& -(U^{0kij}+V^{k0ij})\partial_i A_j. \label{nkkk}
\eqn
The rank of matrix $\pdv[2]{\mathcal L_{(4)}}{\dot A_\mu}{\dot A_\nu}$ is 3, giving rise to a unique primary constraint 
\bqn
\phi_1^{(4)}=\pi^0+(U^{00ij}+V^{ji00})\partial_i A_j+2U^{000i}\partial_iA_0 \approx 0. \nb\\
\label{eq23}
\eqn
The symbol "$\approx$" is called weak equality symbol which implies the equation only holds on the constraint surfaces but not throughout phase space. With the expressions of (\ref{eq4.8}), the canonical Hamiltonian density (\ref{eq17}) also becomes
\begin{widetext}
\bqn
\mathcal H_{(4)}&=&\pi^k\partial_kA_0+\frac{1}{2}{(D^{-1})_{jl}}(\pi^l-N^l)\Big(\pi^j-N^j+({V^{0j}}_{0k}-{V_{k0}}^{j0}){(D^{-1})^{k}}_{i}(\pi^i-N^i)\Big)\nonumber\\
&&+\frac{1}{2}F_{ij}\Big(\frac{1}{2}F^{ij}+\frac{1}{2}W^{ijkl}F_{kl}+2V^{ji0k}\partial_{k}A_{0}+V^{jikl}\partial_{l}A_{k}\Big)\nonumber\\
&&+\qty(\frac{1}{2}U^{ijkl}\partial_i A_j+2U^{0jkl}\partial_j A_0)\partial_k A_l+2U^{00kl}\partial_k A_0\partial_l A_0,
\eqn
\end{widetext}
and the total Hamiltonian density can be written as
\bqn
\mathcal H_{(4)T}=\mathcal H_{(4)}+u^{(4)}\phi_1^{(4)},
\eqn
which gives the total Hamiltonian in the form of
\bqn
H_{(4)T}=\int \mathcal H_{(4)T}(\bvec x)\dd^3x,
\eqn
where $u^{(4)}$ is an arbitrary Lagrangian multiplier. Note that $\partial_t A_0(\bvec x)=\{A_0(\bvec x),H_{(4)T}\}=u^{(4)}(\bvec x)$, because the Poisson bracket between $A_0(\bvec x)$ and $\mathcal H_{(4)}(\bvec y)$ is zero, and between $A_0(\bvec x)$ and other terms in $\phi_1^{(4)}(\bvec y)$, except for $\pi(\bvec y)$, is also zero. This gives meaning to the coefficient $u^{1}(\bvec x)$:  the time derivative of $A_0(\bvec x)$. 

Following the standard Dirac-Bergmann procedure, we then analyze the requirement for the preservation of the primary constraint. Such a requirement is also known as the consistency condition of the primary constraint, which requires that the time derivative of this constraint also vanishes. The consistency condition of the primary constraint, i.e.,
\bqn
\dot \phi_1^{(4)}(\bvec x) =  \{\phi_1^{(4)} (\bvec x), { H}_{(4)T}\} \approx 0,
\eqn
gives rise to a secondary constraint of the system,
\bqn
\phi_2^{(4)}&=&\partial_{k}\pi^k+{M^m}_i(\partial_{m}\pi^i-\partial_{m}N^i) \nb\\
&&+(5U^{00ij}+V^{ji00})\partial_{i}\partial_{j}A_0\nb\\
&& +V^{ji0k}\partial_{k}F_{ij}+2U^{0jkl}\partial_{j}\partial_{k} A_l\approx 0,
\label{eq4.17}
\eqn
where 
\bqn
{M^m}_i&=&\frac{1}{2} \Big[{(D^{-1})_{ij}}+{(D^{-1})_{ji}} \nb\\
&&~~~~ +2({V^{0l}}_{0k}-{V_{k0}}^{l0})(D^{-1})_{li}{(D^{-1})^{k}}_{j}\Big]\nonumber\\
&&\times\qty(3U^{0j0m}+2V^{j00m}+V^{jm00}),
\label{eq2.58}
\eqn
which indicates that the structure of Gauss's law is influenced by the gauge-breaking term.

Similar to the primary constraint, the secondary constraint $\phi_2\approx0$, being a constraint itself, also has a corresponding consistency condition, which is
\bqn
\dot \phi_2^{(4)}(\bvec x) = \{\phi_2^{(4)}(\bvec x) , {H}_{(4)T}\} \approx 0.
\eqn
This condition then leads to 
\bqn\label{eq35}
&&{O_1^{ik}}_j\partial_{i}\partial_{k}(\pi^j-N^j)+{O_2^{ijk}}\partial_{i}\partial_{j}\partial_{k}A_0+O_3^{jilk}\partial_{j}\partial_{i}\partial_{l}A_{k} \nb\\
&& ~~~ +O_4^{klij} \partial_{k} \partial_{l}F_{ij}+T^{ij} \partial_{i}\partial_{j}u^{(4)} \approx 0,
\eqn
where
\bqn
O_3^{jilk}&\equiv&({\delta^j}_m+{M^j}_m)(\frac{1}{2}V^{mikl}+U^{imlk}),\\
O_4^{klij}&\equiv&\frac{1}{2}\qty({\delta^k}_m+{M^k}_m)(\eta^{li}\eta^{mj}+W^{ijlm}+V^{jiml}),\\
{O_1^{ik}}_j&\equiv&\frac{1}{2}\Big[({\delta^i}_m+{M^i}_m)\big(V^{mkl0}+2W^{0lkm} \nb\\
&&~~~~~~~~~~~~~~~~~~~~~~~~  +U^{0lkm}+V^{l0km}\big)\nonumber\\
&&~~ +{M^i}_m(V^{lkm0}+2W^{0mkl}+U^{0mkl}+V^{m0kl}) \nb\\
&& ~~ +2 (V^{li0k}+U^{0ikl})\Big]\nonumber\\
&&\times\Big[{(D^{-1})_{jl}}+{(D^{-1})_{lj}} \nb\\
&&~~~~~~ +2({V^{0{i_1}}}_{0{k_1}}-{V_{{k_1}0}}^{{i_1}0})(D^{-1})_{{i_1}j}{(D^{-1})^{{k_1}}}_{l}\Big],\nb\\
~~~ \\
{O_2^{ijk}}&\equiv& ({\delta^i}_l+{M^i}_l)(V^{lj0k}+2U^{0jkl}) \nb\\
&&+{M^i}_l(U^{0ljk}+V^{l0jk})+2U^{0ijk},
\eqn
and
\bqn
T^{ij} \equiv 6U^{00ij}+(3U^{00jk}+V^{kj00}+2V^{k00j}){M^i}_k. \nb\\
\label{eq2.38}
\eqn
For gauge invariance case, one has $T^{ij}=0$, one can check that the condition (\ref{eq35}) satisfies identically. When the $U(1)$ gauge symmetry of the theory is violated, $T^{ij}$ are in general nonzeros, then (\ref{eq35}) represents a restriction on the Lagrange multiplier $u^{(4)}$. One special case is that $T^{ij}=0$ for some specific combination of the gauge breaking coefficients, for which a new constraint arises, which is
\bqn
\phi_3^{(4)}&=&{O_1^{ik}}_j(\partial_{i}\partial_{k}\pi^j-\partial_{i}\partial_{k}N^j)+{O_2^{ijk}}\partial_{i}\partial_{j}\partial_{k}A_0 \nb\\
&& + O_3^{jilk} \partial_{j}\partial_{i}\partial_{l}A_{k}+O_4^{klij} \partial_{k}\partial_{l}F_{ij}\approx0.
\label{eq69}
\eqn
\textcolor{black}{With this constraint, repeating the above procedure, its consistency condition may produce more constraints under certain conditions. It is worth mentioning that according to Table. 1, the number of  independent coefficients is finite, and the emergence of new constraints will give a set of limiting equations about the coefficients. Just as in the presence of $\phi_3^{(4)}\approx0$, $T^{ij}=0$ gives 9 equations between the coefficients, so that the number of independent coefficients  will be reduced. Detailed analysis will show that each time a new constraint is present, the number of limiting equations for the coefficients will increase quickly as compared to the previous constraint, eventually stopping the generation of possible constraints at a certain step, thus making the Poisson bracket closed and the number of constraints limited. }According to the analysis in \cite{HDOF6, HDOF5}, such constraint may also lead to the unphysical half degree of freedom. However, this requires a very special choice of gauge-breaking coefficients. For simplicity, we will not explore these specific cases in detail in this paper.

Before we go further, we would like to summarize the main results we have gotten so far for the $d=4$ case. For both the gauge invariance case and gauge breaking case with the degenerate condition (\ref{eq4.8}), the theory can have one primary constraint $\phi_1^{(2)}$ and one secondary constraint $\phi_2^{(4)}$.

\subsubsection{Counting the degrees of freedom}

After obtaining all the primary and secondary constraints of the system, let us turn to identify the first- and second-class constraints of the system by analyzing the Poisson bracket of the constraints. In general, the first-class constraints are associated with the gauge symmetry of the theory, they are gauge generators, which generate gauge transformations that don't alter the physical state. The second-class constraints can not generate gauge transformations since the transformation generated by a second-class does not preserve all the constraints, which violates the consistency condition, but are important in the definition of Dirac bracket, which plays a key role in the transition from classical theory to quantum theory \cite{Dirac2}. Specifically, the first-class constraints are those constraints whose Poisson bracket with every constraint vanishes weakly, otherwise are the second-class of constraint.

The Poisson bracket of the primary constraint $\phi_1^{(4)}$ and the secondary constraint $\phi_2^{(4)}$ gives
\bqn
\{\phi_2^{(4)}(y),\phi_1^{(4)}(x)\}&=&(C_1)^{ij}{\partial'}_{i}\partial_{j}\delta(x-y) \nb\\
&& +(C_2)^{ij}{\partial'}_{i}{\partial'}_{j}\delta(x-y).
\label{eq4.21}
\eqn
where $\partial'$ denotes the partial derivative with respect to $y$, and
\bqn
(C_1)^{ij}&=&-\qty(U^{00ji}+(U^{00jk}+V^{kj00}){M^i}_k),\\
(C_2)^{ij}&=&2{M^i}_k(U^{0k0j}+V^{k00j})+5U^{00ij}.
\eqn
It is easy to see that for the gauge invariance case, $C_1^{ij} =0 = C_2^{ij}$, therefore, $\phi_1^{(4)}$ and $\phi_2^{(4)}$ are both first-class constraints. For the case without the gauge symmetry, the gauge breaking coefficients $U^{\mu\nu\rho\sigma}$ and $V^{\mu\nu\rho\sigma}$ are in general nonzero, thus the theory does not possess any first-class constraint. In this case, $\phi_1^{(4)}$ and $\phi_2^{(4)}$ are both second-class constraints. This result is also what we expect since the first-class constraint can only exist when the theory has gauge symmetry, and thus a theory without gauge symmetry should not have first-class constraint.

Getting all first- and second-class constraints, one can count the number of physical degrees of freedom ($N_{\rm DOF}$) by using the following formula \cite{b1},
\bqn
N_{\rm DOF}=\frac{1}{2}(N_{\mathrm {var}}-2\mathrm{NOF}-\mathrm{NOS}),
\eqn
where 
``$N_{\mathrm {var}}$" means the total number of canonical variables, ``NOF" is the number of constraints of the first class, and ``NOS" represents the number of second-class constraints. Thus, for the gauge invariance case, the number of degrees of freedom is
\bqn
N_{\rm DOF} = \frac{1}{2}(8- 2 \times 2) =2,
\eqn
which is the same as that in the standard Maxwell's electrodynamics, while in the case of gauge violation, one has
\bqn
N_{\rm DOF} = \frac{1}{2}(8- 2) =3.
\eqn
Therefore, the violation of the $U(1)$ gauge symmetry induces one additional physical degree of freedom, compared to the standard Maxwell's electrodynamics and the theory with gauge invariance. \textcolor{black}{This additional physical degree of freedom results in a third state of polarization, corresponding to a new particle called longitudinal photon \cite{Ef0}. This represents a new electromagnetic radiation which may alter the radiation spectra of a lot of sources with nonzero temperature \cite{Ef0}. However, its phenomenological effects in both experiments and astrophysical observations are expected to be too small to be detected up to now \cite{Ef0, Ef1,Ef2}.}

\subsection{$d=3$}

In this subsection, let's turn to analyze the constraint structure of the theory with Lorentz/gauge-breaking operators with dimension $d=3$. The Lagrangian density of the considered theory can be written in the form of
\bqn
\mathcal L_{(3)}&=&-\frac{1}{4}F_{\alpha\nu}F^{\alpha\nu}+\frac{1}{2}S^{\mu\nu\rho}A_{\rho}F_{\nu\mu} \nb\\
&& +\frac{1}{2}\epsilon^{\kappa\mu\nu\rho}{(k_{AF})_\kappa}A_{\mu}F_{\nu\rho}.
\label{eq4.20}
\eqn
where
\bqn
{(k_{AF})_\kappa} &=&\frac{1}{3!}\epsilon_{\kappa\mu\nu\rho}{\mathcal{K}}_{(3)}^{(5)\mu\nu\rho},\\
S^{\mu\nu\rho} &=& \mathcal{K}_{(3)}^{(2)\rho\mu\nu}-\mathcal{K}_{(3)}^{(3)\mu\rho\nu}.
\eqn
Note that the terms with ${(k_{AF})_\kappa}$ are gauge-invariant and the terms with $S^{\mu\nu\rho}$ are gauge-breaking. Variation of the action with respect to $A_\mu$, one obtains field equation
\bqn
&& \epsilon^{\kappa\nu\mu\rho} \left({k}_{A F}\right)_\kappa F_{\mu\rho}+\frac{1}{2}S^{\mu\rho\nu}F_{\rho\mu} \nb\\
&&~~~~ +\partial_\mu\qty(F^{\mu\nu}-S^{\nu\mu\rho}A_{\rho})=0.
\eqn

The corresponding conjugate momentum of this theory read,
\bqn
\pi^\mu=&\pdv{\mathcal L_{(3)}}{\dot A_\mu}=-F^{0\mu}+\qty(\epsilon^{\beta\nu0\mu}{k_{AF}}_\beta+S^{\mu0\nu})A_\nu.
\label{eq30}
\eqn
To analyze the constraint structure of the theory, it is convenient to write down the time and spatial components of the above conjugate momentum (\ref{eq30}) as 
\bqn
\pi^0 &=& 0,\\
\pi^k &=& -F^{0k}+\qty(\epsilon^{j\mu0k}{k_{AF}}_{j}+S^{k0\mu})A_\mu.
\eqn
One can conclude that in the specific dimension $d=3$, the gauge-breaking and gauge-invariant Lorentz-breaking electrodynamics have the same form for the conjugate momentum $\pi^0$ of the photon field $A_0$. From these properties, the only primary constraint of the system is
\bqn
\phi_1^{(3)} = \pi^0\approx 0.
\label{eq34}
\eqn
Then, the canonical Hamiltonian density and total Hamiltonian density are given by
\bqn
\mathcal H_{(3)}&=&\pi^k\dot A_k-\mathcal L_{(3)},
\eqn
and
\bqn
(\mathcal H_{(3)})_T&=&\mathcal H_{(3)}+u^{(3)}\phi_1^{(3)}.
\eqn
Here, similar to the case of $d=4$, the coefficient $u^{(3)}$ is also the time derivative of $A_0$. 

Once again, the consistency condition of the primary constraint gives a secondary constraint
\bqn
\phi_2^{(3)}&=&\dot\phi_1^{(3)}\nb\\
&=& \partial_k\pi^k+\frac{1}{2}F_{ik}\qty(\epsilon^{j0ik} \left({k}_{A F}\right)_j  +S^{ki0} ) \nb\\
&&+{S_k}^{00}(\epsilon^{j\mu0k} \left({k}_{A F}\right)_j +S^{k0\mu})A_\mu-\pi^k{S_k}^{00} \nb\\
&\approx & 0.
\label{eq37}
\eqn
Similarly, this means that the form of Gauss's law needs to be modified when considering the presence of the gauge-breaking term. The consistency condition of $\phi_2^{(3)}$ impose a restriction on the $u^{(3)}$:
\bqn
\dot\phi_2^{(3)}&=&{S_l}^{00}S^{l00}u^{(3)} -\big({S_i}^{0k}+{S_i}^{k0}\big)\partial_{k}\pi^i \nonumber\\
&&+{S_k}^{00}\big(2{\epsilon^{jk0}}_i(k_{AF})_j+{S_i}^{0k}-{S^{k0}}_i\big)\pi^i\nonumber\\
&& + {S_i}^{00}\big[2{\epsilon^{l0i}}_k \left({k}_{A F}\right)_l +{S^{i0}}_k -{S_k}^{0i}\big] \nb\\
&&~~~~~~~ \times [{\epsilon^{jn0k}}(k_{AF})_j+S^{k0n}]A_n\nonumber\\
&&+\frac{1}{2}\big(-2S^{j00}{\eta^{ki}}+S^{jik})\partial_{k}F_{ij} \nb\\
&& -\frac{1}{2}(\epsilon^{0 kij} \left({k}_{A F}\right)_0  +S^{jik}){S_k}^{00}F_{ij}\nonumber\\
&&+2{S_l}^{00}(S^{l0k} +S^{lk0 } )\partial_{k}A_0 \nb\\
&& +{S_i}^{00}\big[2{\epsilon^{l0i}}_k \left({k}_{A F}\right)_l +{S^{i0}}_k -{S_k}^{0i}\big]S^{k00}A_0\nonumber\\
&&+\Big[{S_k}^{00}(\epsilon^{0j ik} \left({k}_{A F}\right)_0  +S^{kij }) \nb\\
&&~~~~ + ({S_k}^{0i}+{S_k}^{i0})\qty[{\epsilon^{lj0k}}(k_{AF})_l+S^{k0j}]\Big]\partial_{i}A_j\nonumber\\
&\approx&0.
\label{eq76}
\eqn
For gauge invariance case, all the gauge breaking coefficients $S^{\mu\nu\rho}=0$, and thus the above consistency condition satisfies identically. In this case, the theory only has two constraints, the primary constraint $\phi_1^{(3)}$ and the secondary constraint $\phi_2^{(3)}$.

When the gauge symmetry is violated, the gauge breaking coefficients $S^{\mu\nu\rho}$ are in general nonzeros. For this case, the above consistency condition leads to a specific form of $u^{(3)}$,
\bqn
u^{(3)}&\approx&-\frac{1}{{S_m}^{00}S^{m00}}\\
&&\times\Bigg[{S_i}^{00}\big[2{\epsilon^{l0i}}_k \left({k}_{A F}\right)_l +{S^{i0}}_k -{S_k}^{0i}\big] \nb\\
&&~~~~~~~~~~~  \times [{\epsilon^{jn0k}}(k_{AF})_j+S^{k0n}]A_n\nonumber\\
&&~~~ -\big({S_i}^{0k}+{S_i}^{k0}\big)\partial_{k}\pi^i \nb\\
&&~~~ +{S_k}^{00}\big(2{\epsilon^{jk0}}_i(k_{AF})_j+{S_i}^{0k}-{S^{k0}}_i\big)\pi^i\nonumber\\
&&~~~ +\frac{1}{2}\big(-2S^{j00}{\eta^{ki}}+S^{jik})\partial_{k}F_{ij} \nb\\
&&~~~ -\frac{1}{2}(\epsilon^{0 kij} \left({k}_{A F}\right)_0  +S^{jik}){S_k}^{00}F_{ij}\nonumber\\
&&~~~ +2{S_l}^{00}(S^{l0k} +S^{lk0 } )\partial_{k}A_0 \nb\\
&&~~~ +{S_i}^{00}(2{\epsilon^{l0i}}_k \left({k}_{A F}\right)_l +{S^{i0}}_k -{S_k}^{0i})S^{k00}A_0\nonumber\\
&&~~~ +\Big[{S_k}^{00}(\epsilon^{0j ik} \left({k}_{A F}\right)_0 +S^{kij }) \nb\\
&&~~~~~~~~ +({S_k}^{0i}+{S_k}^{i0})\qty[{\epsilon^{lj0k}}(k_{AF})_l+S^{k0j}]\Big]\partial_{i}A_j\Bigg].\nb\\
\label{eq3.40}
\eqn
This indicates that the theory for this case does not have any additional constraints. Similar to the gauge invariance case, it only has two constraints, one primary constraint $\phi_1^{(3)}$ and one secondary constraint $\phi_2^{(3)}$. Here we would like to mention that, under certain conditions on the gauge breaking coefficients such that $S_{l}^{\;\; 00} S^{l00}=0$, the theory may produce additional constraints. However, this requires a very special choice of gauge-breaking coefficients. For simplicity, we will not explore this specific case in detail in this paper and focus on the case with $S_{l}^{\;\; 00} S^{l00}\neq0$ when gauge symmetry is violated.

Then, let us consider the Poisson bracket of the primary constraint $\phi_1^{(3)}$ and the secondary constraint $\phi_2^{(3)}$, which is
\bqn 
\{\phi_1^{(3)}(\bvec x),\phi_2^{(3)}(\bvec y)\}=-{S_k}^{00}S^{k00}\delta(\bvec x-\bvec y).
\label{eq044}
\eqn
It is obvious that for gauge invariance case, since $S^{\mu\nu\rho}=0$, the above Poisson bracket vanishes and the constraints $\phi_1^{(3)}$ and $\phi_2^{(3)}$ are both first-class. Thus, the number of the physical degrees of freedom is
\bqn
N_{\rm DOF} = \frac{1}{2}(8-2\times 2)=2,
\eqn
which is the same as that in the standard Maxwell's electrodynamics.

For the case with gauge violation, since in general ${S_k}^{00}S^{k00}\neq 0$,  $\phi_1^{(3)},\phi_2^{(3)}$ are both second-class constraints. Thus, the number of the physical degrees of freedom is
\bqn
N_{\rm DOF} = \frac{1}{2}(8-2)=3.
\eqn
Similar to the case with $d=4$, the violation of the $U(1)$ gauge symmetry induces one extra physical degree of freedom, compared to the standard Maxwell's electrodynamics and the case with gauge invariance.

\subsection{$d=2$}

After completing the constraint structure analysis of a specific dimension $d=3$ and $d=4$, let us turn to analyze the structure of $d=2$, which Lagrangian density is
\bqn\label{eqL2}
\mathcal L_{(2)}=-\frac{1}{4}F_{\mu\nu}F^{\mu\nu}+\frac{1}{2}U^{\mu\nu}A_\mu A_\nu,
\eqn
with
\bqn
U^{\mu\nu}&=&2\mathcal{K}_{(2)}^{(1)\mu\nu},
\eqn
which are gauge-breaking terms. Variation of the action with respect to $A_\mu$, one obtains
\bqn
U^{\nu\mu}A_\mu+\partial_\mu F^{\mu\nu}=0
\eqn

The conjugate momenta now read
\bqn
\pi^\mu=\pdv{\mathcal L_{(2)}}{\dot A_\mu}=-F^{0\mu}.
\eqn
One difference from $d=3$ and $d=4$ is that at this point the conjugate momenta $\pi^\mu$ are not affected by gauge-breaking coefficients, $U^{\mu\nu}$, so they are the same as in Maxwell electrodynamics since the time derivative of the photo field $A_\mu$ only appears in the Maxwell term, the first term in (\ref{eqL2}). Therefore, the only primary constraint is 
\bqn
\phi_1^{(2)}=\pi^0\approx0.
\eqn
The canonical Hamiltonian density just differs by one term of $U^{\mu\nu}$ from that in Maxwell electrodynamics:
\bqn
\mathcal H_{(2)}&=&\pi^k\partial_k A_0-\frac{1}{2}\pi^k\pi_k+\frac{1}{4}F_{ik}F^{ik}-\frac{1}{2}U^{\mu\nu}A_\mu A_\nu.\nb\\
\eqn
This gives the total Hamiltonian density:
\bqn
\mathcal H_{(2)T}=\mathcal H_{(2)}+u^{(2)}\phi_1^{(2)}.
\eqn

The influence of gauge-breaking terms are reflected in secondary constraint because it originates from the Poisson bracket between primary constraint and total Hamiltonian which now is gauge-breaking. It reads
\bqn
\phi_2^{(2)}=\partial_k\pi^k+U^{0\mu}A_\mu\approx0.
\eqn
The consistency condition of $\phi_2$ is
\bqn
\dot\phi_2^{(2)}=U^{\mu k}\partial_kA_\mu +U^{0k}\partial_kA_0-U^{0k}\pi_k+u^{(2)}U^{00}\approx 0, \nb\\
\label{eq103}
\eqn
 which provides the constraint on $u^{(2)}$ when $U^{00}\neq0$:
 \bqn
u^{(2)}\approx(U^{00})^{-1}\qty[U^{0k}(\pi_k-\partial_kA_0)-U^{\mu k}\partial_kA_\mu ].
\label{eq2.40}
\eqn
Noticing the absence of $\pi^0$ in $\mathcal H_{(2)}$, we get\textcolor{black}{ $\dot A_0(\bvec x)=\{A_0(\bvec x),H_{(2)T}\}=u^{(2)}(\bvec x)$}. As a consequence, the meaning of $u^{(2)}$ is the time derivative of $A_0$. Combining (\ref{eq2.40}), the time component $A_0$ of the photon field will be determined by the first-order differential equation. 
 
Finally, the Poisson bracket of $\phi_1^{(2)}$ and $\phi_2^{(2)}$ is
\bqn
\{\phi_1^{(2)}(\bvec x),\phi_2^{(2)}(\bvec y)\}=-U^{00}\delta(\bvec x-\bvec y).
\label{eq051}
\eqn
When the gauge breaking coefficients $U^{\mu\nu}$ are nonzeros, $\phi_1^{(2)},\phi_2^{(2)}$ are both second class constraints. Similar to the cases of $d=4$ and $d=3$, we will not explore in detail the case with $U^{00}=0$, which may induce more constraint. Then the number of the physical degree of freedom is
\bqn
N_{\rm DOF} = \frac{1}{2}(8-2) =3.
\eqn
Again, compared to the standard Maxwell electrodynamics, since $d=2$ operators always break the $U(1)$ gauge symmetry, it induces one additional physical degree of freedom.

\section{Map to several specific models \label{se5}}

The Lorentz-violating electrodynamics presented in this paper provide a unifying framework for describing possible violations of Lorentz and $U(1)$ gauge symmetries in the electromagnetic interaction. In this section, we present several specific modified electrodynamics by writing their actions in the form of (\ref{action}) and summarize their Hamiltonian structure from our general analysis. We consider three specific theories, the Lorentz-invariance-Violating (LIV) electrodynamics, the Carroll-Field-Jackiw (CFJ) electrodynamics, and the Proca electrodynamics. The first two theories only break the Lorentz symmetry of the theory, while the last one only breaks $U(1)$ gauge symmetry.

\begin{table*}
\caption{The number of the first-class and second-class constraints, and the number of the physical degrees of freedom for each case with a specific dimension $d=2$, $d=3$, and $d=4$. 
}
\begin{ruledtabular}
\centering
\begin{tabular}{cccccc}
$d$ & gauge invariance & number of first-class constraint & number of first-class constraint   &number of DOF\\
\hline
2 & no & 0 & 2  &3 \\
\hline
\multirow{2}*{3}& yes & 2 & 0 &2\\
&no & 0&2 & 3\\
\hline
\multirow{2}*{4}&yes & 2 & 0 &2\\
&no & 0&2 & 3\\
\end{tabular}
\end{ruledtabular}
\label{tb3.5}
\end{table*}

\subsection{LIV electrodynamics}

We first consider LIV electrodynamics, which is proposed in \cite{HA1,HA2} and the Lagrangian density is given by \cite{HA1,HA2}
\bqn
\mathcal L_{LIV}=-\frac{1}{4}F_{\mu\nu}F^{\mu\nu}-\frac{1}{4}W^{\mu\nu\rho\sigma}F_{\mu\nu}F_{\rho\sigma},
\eqn
which corresponds to the case in our model where the gauge-breaking coefficients $U^{\mu\nu\rho\sigma}$ and $V^{\mu\nu\rho\sigma}$ are set to be zeros for $d=4$. Using the result from (\ref{eq23}) and (\ref{eq4.17}), we obtain the constraint structure in this case:
\bqn
(\phi_1)_{LIV}&=&\pi^0\approx0,\\
(\phi_2)_{LIV}&=&\partial_k\pi^k\approx0.
\eqn
And we need verify that the consistency of $(\phi_2)_{LIV}$ gives no new constraints. Noticing that $T^{ij}=M^{ij}=0$ by using (\ref{eq2.58}) and (\ref{eq2.38}) since $V^{\mu\nu\rho\sigma}=U^{\mu\nu\rho\sigma}=0$, then we only need to check whether $\phi_3^{(4)}$ in (\ref{eq69}) is identically zero throughout the phase space, not just on the constraint surface. In this case, the only possible non-zero coefficients remaining in (\ref{eq69}) are ${O_1^{ik}}_j$ and $O_4^{klij}$. According (\ref{eq48}), as a result, $\phi_3^{(4)}$ now changes to 
\bqn
&&W^{0lik}((D^{-1})_{jl}+(D^{-1})_{lj})\partial_{k}\partial_{i}(\pi^j-N^j) \nb\\
&&~~~~  +\frac{1}{2}(\eta^{li}\eta^{kj}+W^{ijlk})\partial_{k}\partial_{l}F_{ij}.
\label{eq186}
\eqn
Since the last two indices of $W^{ijkl}$ are antisymmetric, and $i,j$ in $\eta^{li}\eta^{kj}\partial_{k}\partial_{l}$ are symmetric, (\ref{eq186}) is equal to 0. This proves that $(\phi_2)_{LIV}$ does not give new constraints. In addition, since the theory has gauge symmetry, both the constraints $(\phi_1)_{LIV}$ and  $(\phi_2)_{LIV}$ are first-class, thus the theory has two degrees of freedom, the same as that in the standard Maxwell electrodynamics. Our result here is consistent with the results in LIV \cite{HA1,HA2}.

\subsection{CFJ electrodynamics}

For CFJ electrodynamics, its Lagrangian density is given by \cite{CFJH}: 
\bqn
\mathcal L_{CFJ}=&-\frac{1}{4}F_{\mu\nu}F^{\mu\nu}+\frac{1}{2}\epsilon^{\kappa\mu\nu\rho}{(k_{AF})_\kappa}A_{\mu}F_{\nu\rho},
\label{eq187}
\eqn
with the caveat that the coefficients  ${(k_{AF})_\kappa}$ differs by a factor of $-\frac{1}{2}$ compared to \cite{CFJH}. The Lagrangian density (\ref{eq187}) can be obtained in our model by setting $S^{\mu\nu\rho}=0$ to zero in $d=3$. Similarly, according to the results in (\ref{eq34}) and (\ref{eq37}), we have two constraints for this theory,
\textcolor{black}{\bqn
&&(\phi_1)_{CFJ}=\pi^0=F^{00}\approx0,\\
&&(\phi_2)_{CFJ}=\partial_k\pi^k+\frac{1}{2}F_{ik}\epsilon^{j0ik} \left({k}_{A F}\right)_j
\approx0,
\eqn
}
where $(\phi_1)_{CFJ}$ is the primary constraint and $(\phi_2)_{CFJ}$ is the secondary constraint.  According to (\ref{eq76}), it can be concluded that in this case, $(\dot\phi_2)_{CFJ}$ is always zero in phase space, which does not generate new constraints. This result confirms \cite{CFJH}, with the additional point to note that the coefficient ${k}_{A F}$ here differs by a factor of $-\frac{1}{2}$ compared to \cite{CFJH}. Note that both the $(\phi_1)_{CFJ}$  and $(\phi_2)_{CFJ}$ are first-class and the number of degrees of freedom is the same as that in the standard Maxwell electrodynamics.

\subsection{The Proca electrodynamics}

When $d=2$, if one sets $U^{\mu\nu}=m^2\eta^{\mu\nu}$, then the Lagrangian density $\mathcal L_{(2)}$ will return to the case of Proca electrodynamics \cite{HA3}:
\bqn
\mathcal L_{Proca}=-\frac{1}{4}F_{\alpha\nu}F^{\alpha\nu}+\frac{1}{2}m^2A_\mu A^\mu,
\eqn
where $m$ is the mass of the photon. Because of this mass term, the Proca electrodynamics breaks the $U(1)$ gauge symmetry of the theory. Replacing the constant coefficients $U^{\mu\nu}$ with the product of metric tensor $\eta^{\mu\nu}$ and $m^2$ is the reason why Proca electrodynamics doesn't break Lorentz symmetry as the product is also a tensor. Same reason as $d=3, 4$, at this point, the theory has two constraints as well, 
\bqn
&&(\phi_1)_{Proca}=\pi^0\approx0,\\
&&(\phi_2)_{Proca}=\partial_k\pi^k+m^2A^0\approx0.
\eqn
And $(\phi_2)_{Proca}$ gives no new constraints. They are both second-class since $\{(\phi_1)_{Proca}(\bvec x),(\phi_2)_{Proca}(\bvec y)\}=m^2\delta(\bvec x-\bvec y)$. Thus this theory propagates three physical degrees of freedom, which is different from the two degrees of freedom in the standard  Maxwell electrodynamics and the cases with gauge invariance. These results are consistent with \cite{HA3}, with the only difference being that their coefficient $m^2$ differs from ours by a factor of $\frac{1}{2}$.

\section{Summary and discussion\label{se6}}

In this paper, we perform an extended analysis of the modified electrodynamics with the violations of both Lorentz symmetry and $U(1)$ gauge symmetry in the framework of the Standard-Model Extension. This represents an extension of the previous construction of the Lorentz-violating electrodynamics with gauge invariance \cite{IS3}. For our purpose, by following the procedure in \cite{IS3}, we construct the quadratic Lagrangian density of electrodynamics by allowing the violations of both the Lorentz and $U(1)$ gauge symmetries. The Lorentz- and gauge-violating effects in the quadratic Lagrangian are represented by the new operators with specific dimension $d \leq 4$. With the constructed quadratic Lagrangian, we calculate in detail the number of independent components of the Lorentz-violating operators at different dimensions with both cases of the gauge invariance and gauge violation cases.

We then perform the Hamiltonian analysis of the general theory by considering Lorentz and gauge-breaking operators with dimension $d \leq 4$ as examples. Specifically, we perform the analysis for $d=4$, $d=3$, and $d=2$, respectively. It is shown that the Lorentz-breaking operators with gauge invariance do not change the classes of the constraints of the theory and have the same number of physical degrees of freedom as that in the standard Maxwell's electrodynamics. When the U(1) gauge symmetry-breaking operators are presented, the theories in general lack first-class constraint and have one additional physical degree of freedom, compared to the standard Maxwell's electrodynamics. The results of the Hamiltonian structure and their corresponding number of degrees of freedom are presented in Table.~\ref{tb3.5}.

Finally, we also map our general analysis to several specific modified electrodynamics, including LIV electrodynamics, CFJ electrodynamics, and Proca electrodynamics. While the former two theories represent two specific examples of Lorentz-violating theory with gauge invariance, the latter one is a theory that breaks $U(1)$ gauge symmetry but still keeps the \textcolor{black}{Lorentz} symmetry. We show that our general results are consistent with the existing Hamiltonian analysis in the literature for these specific examples. 

\section*{Acknowledgments}

This work is supported by the National Key Research and Development Program of China under Grant No. 2020YFC2201503, the Zhejiang Provincial Natural Science Foundation of China under Grant No. LR21A050001 and No. LY20A050002, and the National Natural Science Foundation of China under Grants No.~12275238 and No. 11675143. B.-F. Li is supported by the National Natural Science Foundation of China (NNSFC) with the grant No.~12005186.

\end{document}